\documentclass{article}
\usepackage{amsmath,amssymb,pxfonts,color,graphicx}
\newtheorem{lem}{Lemma}[section]

\newtheorem{thm}{Theorem}[section]
\date{22 March 2019}

\title{Short-term at-the-money asymptotics under
 stochastic volatility models}
 \author{Omar ~El Euch\\
 {\small \'Ecole Polytechnique}\\
 \\
 Masaaki Fukasawa\\
 {\small Graduate School of Engineering Science, Osaka University}\\
 {\small 1-3 Machikaneyama, Toyonaka, Osaka, JAPAN}\\
 {\small fukasawa@sigmath.es.osaka-u.ac.jp }\\
 \\
Jim Gatheral\\
{\small Baruch College, The City University of New York}\\
\\
Mathieu Rosenbaum\\
{\small \'Ecole Polytechnique}
}
\begin{document}
\maketitle
\begin{abstract}
A small-time Edgeworth expansion of the density
 of an asset price is given
 under a general stochastic volatility model,
 from which asymptotic expansions of put option prices and 
 at-the-money implied volatilities follow.
 A limit theorem for at-the-money implied volatility skew and
 curvature
is also given as a corollary. The rough Bergomi model is treated as an example.
 
\end{abstract}
\section{Introduction}
Stochastic volatility models are extensions of the Black-Scholes model
that explain a number of empirical evidences.
The Heston and SABR models among them are popular in financial
practice owing to (semi-)analytic (approximation) formulas for the vanilla option
prices or the option-implied volatilities.
For a practical guide on stochastic volatility modeling,
we refer to \cite{G2006}.
Recently, attracting much attention is a class of stochastic volatility
models where the volatility is driven by a fractional Brownian motion
with the Hurst parameter smaller than $1/2$.
This is due to their consistency to a power law for the term structure of
the implied volatility skew that has been empirically recognized;
see \cite{Alos, BFG, FZ, F2011, F2017,  GS, GJR, GJR2}.
The small Hurst parameter
implies in particular  that the volatility path is rougher than a Brownian
motion and so, this class is often referred as the rough
volatility models.
Since the models do not admit of explicit expressions for option prices
or implied volatilities, the above mentioned consistency has been
discussed through asymptotic analyses.

The aim of this paper is to provide a general framework under which the
short-term asymptotics of the at-the-money implied volatility is
studied.
 Here, by short-term asymptotics, we mean the asymptotics
with time-to-maturity $\theta \to 0$. 
By at-the-money, we mean a regime of log-moneyness $k=
O(\sqrt{\theta})$.
The framework is for a general continuous stochastic volatility model.
The rough Bergomi model introduced by \cite{BFG} is treated as an example.
The asymptotic expansion of the at-the-money implied volatility is given
up to the second-order.

The first-order expansion was already given in
\cite{F2017} by a different method.
 For the SABR model, Osajima \cite{Osa} gave the second-order 
expansion based on the
Watanabe-Yoshida theory; see e.g.,~\cite{KT, Y1992}.
 For a Markov stochastic volatility model with jumps,
Medvedev and Scaillet~\cite{MS0,MS} derived 
the expansion  by a formal computation.
For the Markov diffusion case, 
Pagliarani and Pascucci~\cite{PP} proved the validity of the Taylor
 expansion.
An expansion of the  at-the-money implied volatility 
skew is derived under a L\'evy jump model with Markov
stochastic volatility
 by Figueroa-L\'opez and \'Olafsson~\cite{FO}.
Beside these results for the at-the-money regime, considering
near-the-money, that is, a moderate deviation regime,
Friz et al.~\cite{FGP} derived the asymptotic skew and curvature of the
implied volatility 
by assuming the asymptotic behavior of the density function of the
underlying asset price.
 Recently, 
Bayer et al.~\cite{Friz} has extended the moderate deviation analysis
 to a rough volatility model.

In this paper, we introduce a novel approach based on the
conditional Gaussianity
of a  continuous stochastic volatility model to prove the validity of a second
order density expansion, from which 
follow expansions of the option prices and the implied volatility as well as the asymptotic skew and curvature formula.
 In contrast to \cite{KT, Osa, Y1992}, we do not rely on the Malliavin
 calculus, which enables us to treat effectively the rough volatility models.
In contrast to the elementary method of \cite{F2017}, our approach can
be extended to higher-order expansions without any
 additional theoretical difficulty.
We choose the square root of the forward variance, that is, the fair strike of a variance swap, as the leading term of our asymptotic expansion, while a recent work~\cite{AlSh} studies the difference between the implied volatility and the fair strike of a volatility swap in terms of the Malliavin derivatives.
 
The paper is organized as follows. In Section~2, we describe assumptions and general results.
In Section~3, we give the proofs of the general results.
In Section~4, we treat regular stochastic volatility models.
In Section~5, we show that the rough Bergomi model fits into the framework as well
and
compute the coefficients of the expansion for this particular model.

 \section{Framework}
 \subsection{Assumptions}
Let $(\Omega,\mathcal{F},Q)$ be a 
probability space equipped with a filtration
$\{\mathcal{F}_t ; t \geq 0\}$ satisfying the usual assumptions.
A log price process $Z$ is assumed to follow
\begin{equation*}
\mathrm{d}Z_t = r\mathrm{d}t - \frac{1}{2} v_t \mathrm{d}t
 + \sqrt{v}_t\mathrm{d}B_t,
 \end{equation*}
 where $r\in\mathbb{R}$ stands for an interest rate and $v$ is a
 positive continuous process adapted to
 a smaller filtration $\{\mathcal{G}_t; t \geq 0\}$, of which the square root is called the volatility
 of $Z$. The Brownian motion $B$ is decomposed as
 \begin{equation*}
 \mathrm{d}B_t = \rho_t \mathrm{d}W_t +
 \sqrt{1-\rho^2_t}\mathrm{d}W^\prime_t,
 \end{equation*}
 where $W^\prime$ is an $\{\mathcal{F}_t\}$-Brownian motion independent of
 $\mathcal{G}_t$ for all $t\geq 0$,
 $W$ is a $\{\mathcal{G}_t\}$-Brownian motion and $\rho$ is a progressively
 measurable processes with respect to $\{\mathcal{G}_t\}$ and taking
 values in $[-1,1]$.
A typical situation for stochastic volatility models, including the Heston, SABR and rough Bergomi models,  is that $(W,W^\prime)$ is a two dimensional
$\{\mathcal{F}_t\}$-Brownian motion and $\{\mathcal{G}_t \}$ is the
filtration generated by $W$, that is,
\begin{equation*}
\mathcal{G}_t = \mathcal{N} \vee \sigma(W_s; s \leq t),
\end{equation*}
where $\mathcal{N}$ is the null sets of $\mathcal{F}$.
Denote by $\|\cdot \|_p$ the $L^p$ norm under  $Q$.
Our key assumption is the following:
for any $p > 0$, 
\begin{equation}\label{smooth}
\sup_{\theta \in (0,1)}\left\|
\frac{1}{\theta}\int_0^\theta v_t \mathrm{d}t\right\|_p < \infty, \ \
\sup_{\theta \in (0,1)}\left\|\left\{ \frac{1}{\theta}
\int_0^\theta v_t(1-\rho_t^2) \mathrm{d}t \right\}^{-1} \right\|_p
< \infty.
\end{equation}
This is satisfied by standard stochastic volatility models (with correlation parameter $|\rho| < 1$) but not by local volatility models that correspond to
 $\rho \equiv 1$.\\

An arbitrage-free price $p(K,\theta)$ of a put option at time $0$ with strike $K>0$ and
maturity $\theta>0$ is given by
\begin{equation*}
p(K,\theta) = e^{-r\theta}E[(K- \exp(Z_\theta))_+] =
e^{-r\theta}\int_0^K Q(\log x \geq Z_\theta ) \mathrm{d}x.
\end{equation*}
The forward variance curve $v_0(t)$ at time $0$ is defined by
$v_0(t) = E[v_t]$.
Changing variable as
\begin{equation*}
x = F\exp\left(\zeta \sigma_0(\theta) \right), \ \ 
F = \exp(r\theta + Z_0),
\end{equation*}
where
\begin{equation*}
\sigma_0(\theta) = \sqrt{\int_0^\theta v_0(t)\mathrm{d}t},
\end{equation*}
we have
\begin{equation*}
\frac{p(Fe^{z\sigma_0(\theta)},\theta)}
{F\sigma_0(\theta)}
=e^{-r\theta}\int_{-\infty}^{z} Q\left( \zeta \geq X_\theta \right) e^{\sigma_0(\theta)\zeta}\mathrm{d}\zeta,
\end{equation*}
where
\begin{equation*}
X_\theta = -\frac{1}{2\sigma_0(\theta)}\langle M
\rangle_\theta + \frac{1}{\sigma_0(\theta)} M_\theta, \ \ 
M_\theta = \int_0^\theta \sqrt{v_t} \mathrm{d}B_t, \ \
\langle M \rangle_\theta = \int_0^\theta v_t \mathrm{d}t.
\end{equation*}
Based on this expression, the asymptotic behavior of put option prices is studied through the asymptotic distribution of $X_\theta$.
From the martingale central limit theorem\footnote{
The martingale central limit theorem for one-dimensional continuous
local martingales
is proved as follows. Let $M^n$ be a continuous local
martingale with $\langle M^n \rangle_1 \to 1$ in probability. 
By the Dambis-Dubins-Schwarz theorem, $M^n = W^n_{\langle M^n
\rangle}$ for a Brownian motion $W^n$. Since $(W^n,\langle M^n
\rangle_1) \to (W,1)$ in law, by the continuous mapping theorem, we conclude
$M^n_1 \to W_1$ in law.}, 
it is not difficult to see that  $X_\theta$ converges in law to
the standard normal distribution
as $\theta \to 0$.
To determine higher-order asymptotic distribution,  
we assume the following structure:
there exists a family of random vectors
\begin{equation*}
\left\{ (M^{(0)}_\theta, M^{(1)}_\theta, M^{(2)}_\theta,
M^{(3)}_\theta) ; \theta \in (0,1) \right\}
\end{equation*}
such that 
\begin{enumerate}
\item the law of $M^{(0)}_\theta$ is standard normal for all $\theta
>0$,
\item 
\begin{equation}\label{moments}
\sup_{\theta \in (0,1)} \|M^{(i)}_\theta\|_p < \infty, \ \ i=1,2,3
\end{equation}
for all $p > 0$ and 
\item 
for some $H \in (0,1/2]$ and $\epsilon \in (0,H)$,
\begin{equation}\label{stex}
\begin{split}
&\lim_{\theta \to 0} \theta^{-2H - 2\epsilon}\left\|
\frac{M_\theta}{\sigma_0(\theta)} - M^{(0)}_\theta - \theta^H
M^{(1)}_\theta - \theta^{2H} M^{(2)}_\theta \right\|_{1+\epsilon}
=0, \\
&
\lim_{\theta \to 0} \theta^{- H-2 \epsilon}\left\| 
\frac{\langle M \rangle_\theta}{\sigma_0(\theta)^2} - 1 - \theta^H M^{(3)}_\theta\right\|_{1+\epsilon}
=0.
\end{split}
\end{equation}
\end{enumerate}
Further, we assume the existence of the derivatives
\begin{equation}\label{ab}
\begin{split}
&a^{(i)}_\theta(x) = \frac{\mathrm{d}}{\mathrm{d}x}\left\{E[M^{(i)}_\theta |
M^{(0)}_\theta=x]\phi(x)\right\}, \ \ i= 1,2,3, \\
&b_\theta(x) = \frac{\mathrm{d}^2}{\mathrm{d}x^2}\left\{ E[M^{(1)}_\theta |
M^{(0)}_\theta=x] \phi(x) \right\} \\
&c_\theta(x) = \frac{\mathrm{d}^2}{\mathrm{d}x^2}\left\{ E[|M^{(1)}_\theta|^2 |
M^{(0)}_\theta=x] \phi(x) \right\}
\end{split}
\end{equation}
in the Schwartz space (i.e., the space of the rapidly decreasing smooth
functions), where $\phi$ is the standard normal density. 

As will be discussed in Section~4, regular stochastic volatility models satisfy these assumptions with $H = 1/2$, where (\ref{stex}) is a consequence of the It\^o-Taylor expansion.
In Section~5, we see that the rough Bergomi model, where the volatility is driven by a fractional Brownian motion, 
satisfies these assumptions with $H$ being the Hurst parameter of the fractional Brownian motion.

\subsection{General results}
The fundamental result in this paper is the following.
\begin{thm}
 \label{thm:denex}
   The law of  $X_\theta$ admits a density  $p_\theta$,  and
   for any $\alpha \in \mathbb{N} \cup \{0\}$, 
   \begin{equation}\label{denex}
   \sup_{x \in \mathbb{R}}
   (1+x^2)^{\alpha}|p_\theta(x) - q_\theta(x)| = o(\theta^{2H})
   \end{equation}
   as $\theta \to 0$, where
  \begin{equation}\label{qth2}
\begin{split}
  q_\theta(x) =  &\phi\left(x + \frac{\sigma_0(\theta)}{2}\right) - \theta^H \left(a^{(1)}_\theta\left(x + \frac{\sigma_0(\theta)}{2}\right)-  \frac{\sigma_0(\theta)}{2} a^{(3)}_\theta\left(x + \frac{\sigma_0(\theta)}{2}\right)\right) \\
& -
  \theta^{2H}\left( a^{(2)}_\theta(x) -  \frac{1}{2}c_\theta(x) \right).
\end{split}
  \end{equation}
\end{thm}
The proof is given in Section~3.2. In order to derive a neat asymptotic expansion of the put option prices, we introduce an additional assumption which is satisfied by the models in Sections~4 and 5.

\begin{thm}\label{thm:prex}
Suppose we have (\ref{denex}) with $q_\theta$ of the form
  \begin{equation}\label{thm32form}
\begin{split}
  q_\theta(x)   =  & \phi\left(x + \frac{\sigma_0(\theta)}{2}\right) \left\{ 1 + 
 \kappa_3(\theta) \left(H_3 \left(x + \frac{\sigma_0(\theta)}{2}\right)- \sigma_0(\theta) H_2 \left(x + \frac{\sigma_0(\theta)}{2}\right)\right)\theta^H 
\right\} \\
& + \phi(x) \left(\kappa_4(\theta)  H_4(x) +
  \frac{\kappa_3(\theta)^2}{2}H_6(x)\right) \theta^{2H}
  \end{split}\end{equation}
with bounded functions $\kappa_3(\theta)$  and $\kappa_4(\theta)$ of $\theta$, 
 where $H_k$ is the $k$th Hermite polynomial:
  \begin{equation*}
  H_1(x) = x,\ \ H_2(x) = x^2-1, \ \ 
  H_3(x) = x^3-3x, \ \ H_4(x) = x^4 - 6x^2 + 3,  \dots
  \end{equation*}
  Then, for any $z_0 \in\mathbb{R}$,
  \begin{equation*}
  \begin{split}
 \frac{p(Fe^{\sigma_0(\theta)z},\theta)}
{Fe^{-r\theta}\sigma_0(\theta)}
 & = 
\frac{1}{\sigma_0(\theta)}\left(\Phi\left(z + \frac{\sigma_0(\theta)}{2}\right)e^{\sigma_0(\theta)z}-\Phi\left(z-\frac{\sigma_0(\theta)}{2}\right) \right) \\
 & \hspace{0.5cm} + \kappa_3(\theta) \phi\left(z+\frac{\sigma_0(\theta)}{2}\right) 
  H_1\left(z+\frac{\sigma_0(\theta)}{2}\right)e^{\sigma_0(\theta)z} \theta^H 
    \\
& \hspace{0.5cm}
 + \phi(z)\left(\kappa_4(\theta)  H_2(z) +
  \frac{\kappa_3(\theta)^2}{2}H_4(z)\right) \theta^{2H}+ o(\theta^{2H})
  \end{split}
  \end{equation*}
  uniformly in $z \leq z_0$.
  \end{thm}
The proof is given in Section~3.3. Under the same assumption,  an asymptotic expansion of the Black-Scholes implied volatility follows.
  Denote by $p_{\rm{BS}}(K,\theta,\sigma)$ the put option price
  with strike price $K$ and maturity $\theta$ under the Black-Scholes
  model with volatility parameter $\sigma > 0$.
  Given a put option price $p(K,\theta)$, $K = Fe^k$,
the Black-Scholes implied volatility $\sigma_{\rm{BS}}(k,\theta)$ is defined through
  \begin{equation*}
  p_{\rm{BS}}(K,\theta,\sigma_{\rm{BS}}(k,\theta)) = p(K,\theta).
  \end{equation*}
The at-the-money implied volatility skew and curvature
are defined respectively as the first and the second derivatives in $k$ of the
Black-Scholes implied volatility   at $k = 0$.
   The skew behavior is especially important in order to argue the
   consistency of a  model to the empirically observed power law.

  \begin{thm} \label{thm:ivex}
Suppose we have (\ref{denex}) with $q_\theta$ of the form (\ref{thm32form}).
Then, 
   for any $z \in \mathbb{R}$,
  \begin{equation*}
\begin{split}
   & \sigma_{\rm{BS}}(\sqrt{\theta}z,\theta) \\ & =
  \kappa_2\left\{
  1 + \kappa_3 \left(
   \frac{z}{\kappa_2} + \frac{\kappa_2\sqrt{\theta}}{2}  \right) \theta^H
  + \left(  \frac{3 \kappa_3^2}{2}-\kappa_4 + 
  (\kappa_4 - 3\kappa_3^2) \frac{z^2} {\kappa_2^2}
   \right) \theta^{2H} \right\}
  + o(\theta^{2H}),
\end{split}
  \end{equation*}
   where $\kappa_2 = \kappa_2(\theta) = \sigma_0(\theta)/\sqrt{\theta}$,
  $\kappa_3 = \kappa_3(\theta)$ and $\kappa_4 = \kappa_4(\theta)$.
 
    \end{thm}
  \begin{thm}\label{thm:asbe}
Suppose we have (\ref{denex}) with $q_\theta$ of the form (\ref{thm32form}).
Then, 
      \begin{equation*}
      \begin{split}
 &     \partial_k \sigma_{\rm{BS}}(0,\theta)
      = \kappa_3(\theta)\theta^{H-1/2} + o(\theta^{2H-1/2}), \\
&            \partial_k^2 \sigma_{\rm{BS}}(0,\theta)
	    = 2\frac{\kappa_4(\theta) - 3\kappa_3(\theta)^2
  }{\kappa_2(\theta)} \theta^{2H-1} + o(\theta^{2H-1}).
     \end{split} \end{equation*}
      \end{thm}
The proofs are given in Sections 3.4 and 3.5 respectively.
\\

\noindent
{\bf Remark: }
The above asymptotic estimates are not uniform in $H$; the assumed stochastic expansion (\ref{stex}) is not uniform in $H$ and so, there seems no hope to have uniformity. We would have uniformity in $H \in [H_0, 1/2]$ for some $H_0 > 0$ if we could strengthen the condition (\ref{stex}) to uniform convergence on $[H_0,1/2]$. It seems impossible to argue the uniformity in $H \in (0,1/2]$ because Lemma 3.2 below requires  some $\epsilon$-$\delta$ argument depending on $H$.

\section{Proofs}
\subsection{Characteristic function expansion}
Here we give an asymptotic expansion of the characteristic function of
$X_\theta$. Let
\begin{equation*}
Y_\theta = M^{(0)}_\theta + \theta^H M^{(1)}_\theta +
\theta^{2H}M^{(2)}_\theta
- \frac{\sigma_0(\theta)}{2} \left( 1 + \theta^H M^{(3)}_\theta\right).
\end{equation*}
\begin{lem}\label{lem31} 
 Let $H \in (0,1/2]$ and
$\epsilon \in (0,H)$ be constants under
 which (\ref{stex}) holds. Then, 
 for any $\alpha \in \mathbb{N}\cup \{0\}$,
\begin{equation*}
\sup_{|u|\leq \theta^{-\epsilon}} |E[X_\theta^\alpha e^{iuX_\theta}]
- E[Y_\theta^\alpha e^{iuY_{\theta}}]| = o(\theta^{2H + \epsilon}).
\end{equation*}
\end{lem}
{\it Proof: }
Since $|e^{ix}-1|\leq |x|$, we have
\begin{equation*}
\begin{split}
|E[X_\theta^\alpha e^{iuX_\theta}]
- E[Y_\theta^\alpha e^{iuY_{\theta}}]|
&\leq |E[(X^\alpha_\theta -Y^\alpha_\theta )e^{iu X_\theta}]|
+ |E[Y^\alpha_\theta e^{iu Y_\theta}(e^{iu(X_\theta-Y_\theta)}-1)]|\\
&\leq E[|X_\theta^\alpha - Y_\theta^\alpha|] + u
E[|Y_\theta|^\alpha|X_\theta-Y_\theta|] 
\end{split}
\end{equation*}
By (\ref{smooth}) and (\ref{moments}) respectively, $X_\theta$ and $Y_\theta$ 
have moments of any order. Therefore by the H\"older inequality,
\begin{equation*}
 E[|Y_\theta|^\alpha|X_\theta-Y_\theta|] \leq C_1(\alpha,\epsilon)
\|X_\theta - Y_\theta \|_{1+\epsilon}
\end{equation*}
for a constant $C_1(\alpha,\epsilon) > 0$.
Since $X^\alpha_\theta - Y^\alpha_\theta =
(X_\theta-Y_\theta)\sum_{\beta=0}^{\alpha-1} (-1)^\beta
X_\theta^{\alpha-1-\beta}Y_\theta^\beta$,
the H\"older inequality gives also 
\begin{equation*}
E[|X_\theta^\alpha - Y_\theta^\alpha|] \leq 
C_2(\alpha,\epsilon)\|X_\theta - Y_\theta \|_{1+\epsilon}
 \end{equation*}
 for a constant $C_2(\alpha,\epsilon) > 0$.
 Since $\sigma_0(\theta) = O(\theta^{1/2})$, we have 
$\|X_\theta - Y_\theta \|_{1+\epsilon} = o(\theta^{2H+2\epsilon})$ 
by (\ref{stex}),
from which the result follows. \hfill////

 \begin{lem}
Let $H$ and $\epsilon $ be as in Lemme~\ref{lem31}.
Then,  for any $\delta \in [0,(H-\epsilon)/3)$,
  \begin{equation*}
\sup_{|u| \leq \theta^{-\delta}} \left| E[Y_\theta^\alpha e^{iuY_\theta}]
 - E\left[e^{iuM^{(0)}_\theta}\left((M^{(0)}_\theta)^\alpha + A(\alpha,u,M^{(0)}_\theta) +
  B(\alpha,u,M^{(0)}_\theta)\right)\right] \right|= o(\theta^{2H + \epsilon}),
 \end{equation*}
where
 \begin{equation*}
 \begin{split}
 A_\theta(\alpha,u,x) = &  \left(iux^\alpha + \alpha
  x^{\alpha-1}\right)( E[Y_\theta| M^{(0)}_\theta = x]-x),\\
   B_\theta(\alpha,u,x) =  &\left( -\frac{u^2}{2}x^{\alpha} + iu \alpha x^{\alpha-1} +
  \frac{\alpha(\alpha-1)}{2}x^{\alpha-2}\right) \\ 
& \times
  \left( \theta^{2H}E[|M^{(1)}_\theta|^2 | M^{(0)}_\theta = x] 
-\sigma_0(\theta)\theta^HE[M^{(1)}_\theta | M^{(0)}_\theta = x] +
\frac{\sigma_0(\theta)^2}{4}   \right).
  \end{split}
  \end{equation*}
  \end{lem}
  {\it Proof: }
  This follows from the fact that
  \begin{equation*}
  \left| e^{ix}-1 - ix + \frac{x^2}{2} \right| \leq \frac{|x|^3}{6}
  \end{equation*}
  for all $x \in \mathbb{R}$. 
 Indeed, this implies that
  \begin{equation*}
\sup_{|u| \leq \theta^{-\delta}} \left| E[Y_\theta^\alpha e^{iuY_\theta}]
 - E\left[Y_\theta^\alpha e^{iuM^{(0)}_\theta}\left(1 + iu(Y_\theta -
					       M^{(0)}_\theta) -
					       \frac{u^2}{2}(Y_\theta - M^{(0)}_\theta)^2\right)\right]
 \right|= o(\theta^{2H + \epsilon}).
 \end{equation*}
Expand $Y^\alpha_\theta = (M^{(0)}_\theta)^\alpha + \alpha
(M^{(0)}_\theta)^{\alpha-1}(Y_\theta - M^{(0)}_\theta) + \dots $ and
take the conditional expectation given $M^{(0)}_\theta $ to obtain
the result.
\hfill////

  \begin{lem}
  Define $\bar{q}_\theta(x)$ by
  \begin{equation}\label{qth}
\begin{split}
  \bar{q}_\theta(x) =  &\phi(x) - \theta^H a^{(1)}_\theta(x) -
  \theta^{2H}a^{(2)}_\theta(x) - \frac{\sigma_0(\theta)}{2}(x\phi(x) -
  \theta^Ha^{(3)}_\theta(x)) \\ &+ \frac{\theta^{2H}}{2}c_\theta(x)
- \frac{\theta^H  \sigma_0(\theta)}{2} b_\theta(x) + \frac{\sigma_0(\theta)^2}{8}(x^2-1)\phi(x),
\end{split}
  \end{equation}
where $a^{(i)}_\theta$, $b_\theta$  and $c_\theta$ are defined by (\ref{ab}).
  Then,
  \begin{equation*}
  \int_\mathbb{R} e^{iux} x^\alpha \bar{q}_\theta(x) \mathrm{d}x
  = E\left[e^{iuM^{(0)}_\theta}\left((M^{(0)}_\theta)^\alpha + A(\alpha,u,M^{(0)}_\theta) +
  B(\alpha,u,M^{(0)}_\theta)\right)\right].
  \end{equation*}
  \end{lem}
  {\it Proof: }
Since the density of $M^{(0)}_\theta$ is $\phi$ by the assumption, 
this simply follows from integration by parts. \hfill////
  \subsection{Density expansion}
  Here we derive an asymptotic expansion of the density of $X_\theta$.
  \begin{lem} \label{lem:smooth}
There exists a density of $X_\theta$ and
  for any $\alpha, j \in \mathbb{N} \cup \{0\}$,
  \begin{equation*}
\sup_{\theta \in (0,1)}  \int |u|^j |E[X_\theta^\alpha e^{iuX_\theta}]| \mathrm{d}u < \infty
  \end{equation*}
  \end{lem}
  {\it Proof: }
  Note that the distribution of $X_\theta$ is Gaussian conditionally on
  $\mathcal{G}_\theta$,
with conditional mean 
\begin{equation*}
U_\theta:=  -\frac{1}{2\sigma_0(\theta)}\langle M
\rangle_\theta + \frac{1}{\sigma_0(\theta)} \int_0^\theta \sqrt{v_t}\rho_t\mathrm{d}W_t
\end{equation*}
and conditional variance
\begin{equation*}
V_\theta :=  \frac{1}{\sigma_0(\theta)^2} \int_0^\theta v_t(1-\rho_t^2)\mathrm{d}t.
\end{equation*}
Therefore, for any bounded continuous function $f$, we have
\begin{equation*}
 E[f(X_\theta)]
= E[E[f(X_\theta)|\mathcal{G}_\theta]]
= E[ \int f(x)\phi(x,U_\theta,V_\theta) \mathrm{d}x],
\end{equation*}
where $\phi(\cdot,u,v)$ is the density of the normal distribution with
mean $u$ and variance $v$. This means that  
$X_\theta$ 
admits a density 
\begin{equation*}
 p_\theta(x) = E[\phi(x,U_\theta,V_\theta)].
\end{equation*}
Furthermore, the density function is in the Schwartz space $\mathcal{S}$
and each Schwartz semi-norm is uniformly bounded  in $\theta$  by (\ref{smooth}).
  Therefore,
  \begin{equation*}
  \begin{split}
  \sup_{\theta \in (0,1)} \int |u|^j |E[X_\theta^\alpha e^{iuX_\theta}]| \mathrm{d}u
   &= \sup_{\theta \in (0,1)} \int\left| \int u^jx^\alpha e^{iux}p_\theta(x)\mathrm{d}x \right|
   \mathrm{d}u
   \\ &= \sup_{\theta \in (0,1)} \int \left|
   \int e^{iux} \partial_x^j(x^\alpha p_\theta (x)) \mathrm{d}x
   \right| \mathrm{d}u < \infty
   \end{split}
   \end{equation*}
   since the Fourier transform is a continuous linear mapping from $\mathcal{S}$ to $\mathcal{S}$.
   \hfill////
\\

\noindent
   {\it Proof of Theorem~\ref{thm:denex}: }
   As seen in the proof of Lemma~\ref{lem:smooth}, the density
   $p_\theta$ exists in the Schwartz space.
Note that for a function $f$ in the Schwartz space,
by Taylor's theorem,
\begin{equation*}
\begin{split}
\left| f(x+a) - f(x) -f^\prime(x)a - f^{\prime\prime}(x)\frac{a^2}{2}
 \right|
 & \leq \frac{a^3}{2} \sup_{|b|\leq |a|}
 |f^{\prime\prime\prime}(x+b)|
\\ &\leq \frac{a^3}{2} \sup_{|b|\leq |a| } \frac{1}{(1 + (x+b)^2)^\alpha}
\sup_{y \in \mathbb{R}}
 (1+y^2)^\alpha |f^{\prime\prime \prime}(y)|
\end{split}
\end{equation*}
and so,
\begin{equation*}
\sup_{x \in \mathbb{R}}
(1+x^2)^\alpha
\left| f(x+a) - f(x) -f^\prime(x)a - f^{\prime\prime}(x)\frac{a^2}{2} \right|
 = O(a^3).
\end{equation*}
This gives
  \begin{equation*}
   \sup_{x \in \mathbb{R}} (1+x^2)^{\alpha}
   |q_\theta(x) - \bar{q}_\theta(x)|
   = O(\theta^{1+H}) = o(\theta^{2H}),
\end{equation*}
where $\bar{q}_\theta$ is given by (\ref{qth}).
By the Fourier identity,
   \begin{equation*}
    (1+x^2)^{\alpha}
   (p_\theta(x) - \bar{q}_\theta(x))
   = \frac{1}{2\pi}
   \int
\int e^{iuy} (1+y^2)^\alpha(p_\theta(y)-\bar{q}_\theta(y))\mathrm{d}y e^{-i ux}\mathrm{d}u
   \end{equation*}
   Combining the lemmas in the previous section, taking $\delta \in (0,
   \min\{\epsilon, (H-\epsilon)/3\})$,
   we have
   \begin{equation*}
    \int_{|u|\leq \theta^{-\delta}} \left|
\int e^{iuy} (1+y^2)^\alpha(p_\theta(y)-\bar{q}_\theta(y))\mathrm{d}y
   \right| \mathrm{d}u = o(\theta^{2H}).
   \end{equation*}
   On the other hand,
   \begin{equation*}
   \begin{split}
   \int_{|u| \geq \theta^{-\delta}} \left| \int e^{iuy}
   (1+y^2)^\alpha p_\theta(y) \mathrm{d}y  \right| \mathrm{d}u
   &\leq \theta^{j \delta}
   \int_{|u| \geq \theta^{-\delta}} |u|^j| E[(1+X_\theta^2)^\alpha
   e^{iuX_\theta}]|\mathrm{d}u
   \\ & = O(\theta^{j\delta})
   \end{split}
   \end{equation*}
   for any $j \in \mathbb{N}$ by Lemma~\ref{lem:smooth}.
   The remainder
   \begin{equation*}
   \int_{|u| \geq \theta^{-\delta}} \left| \int e^{iuy}
   (1+y^2)^\alpha \bar{q}_\theta(y) \mathrm{d}y  \right| \mathrm{d}u
   \end{equation*}
  is handled in the same manner. \hfill////

    \subsection{Put option price  expansion}
    Here we consider put option prices.
    Denote by $p_\theta$ the density of $X_\theta$ as before and 
   consider a normalized  put option price
\begin{equation*}
\frac{p(Fe^{\sigma_0(\theta)z},\theta)}
{F\sigma_0(\theta)}
=e^{-r\theta}\int_{-\infty}^{z}\int_{-\infty}^\zeta p_\theta(x)
  \mathrm{d}x
  e^{\sigma_0(\theta)\zeta}\mathrm{d}\zeta.
\end{equation*}
\begin{lem}
Let $q_\theta(x)$, $\theta>0$ be a family of functions on $\mathbb{R}$ (not necessarily the one given by (\ref{qth})). If
\begin{equation*}
\sup_{x \in \mathbb{R}} (1+x^2)^{\alpha} |p_\theta(x)-q_\theta(x)| =
o(\theta^\beta)
\end{equation*}
for some $\alpha > 5/4$ and $\beta >0$,
then for any $z_0 \in \mathbb{R}$,
\begin{equation*}
\frac{p(Fe^{\sigma_0(\theta)z},\theta)}
{F\sigma_0(\theta)}
=e^{-r\theta}\int_{-\infty}^{z}\int_{-\infty}^\zeta q_\theta(x)
\mathrm{d}x e^{\sigma_0(\theta)\zeta}\mathrm{d}\zeta + o(\theta^\beta)
\end{equation*}
uniformly in $z \leq z_0$.
\end{lem}
{\it Proof: }
By the Cauchy-Schwarz inequality,
\begin{equation*}
\begin{split}
&e^{-r\theta}\int_{-\infty}^{z}\int_{-\infty}^\zeta |p_\theta(x) - q_\theta(x)|
\mathrm{d}z e^{\sigma_0(\theta)\zeta}\mathrm{d}\zeta
\\ &\leq
e^{-r\theta}
\int_{-\infty}^z \sqrt{ \int_{-\infty}^{\zeta}
\frac{\mathrm{d}x}{(1+x^2)^{2\alpha-1}}}
\sqrt{\int_{-\infty}^{\zeta} (1+x^2)^{2\alpha-1}|p_\theta(x)-q_\theta(x)|^2
\mathrm{d}z } e^{\sigma_0(\theta)\zeta}\mathrm{d}\zeta
\\
& \leq \sqrt{\pi} e^{-r\theta + \sigma_0(\theta)z}
\sup_{x \in \mathbb{R}} (1+x^2)^{\alpha} |p_\theta(x)-q_\theta(x)|
\int_{-\infty}^z \sqrt{ \int_{-\infty}^{\zeta}
\frac{\mathrm{d}x}{(1+x^2)^{2\alpha-1}}} \mathrm{d}\zeta,
\end{split}
\end{equation*}
which is $o(\theta^\beta)$ if $\alpha > 5/4$. \hfill////
\\

\noindent
  {\it Proof  of Theorem~\ref{thm:prex}: }
  This is a direct consequence of the previous lemma. For example,
\begin{equation*}
\frac{\mathrm{d}}{\mathrm{d}z} \left\{e^{-\sigma_0(\theta)z} \frac{\mathrm{d}}{\mathrm{d}z} \left\{
\frac{1}{\sigma_0(\theta)}\left(\Phi\left(z + \frac{\sigma_0(\theta)}{2}\right)e^{\sigma_0(\theta)z}-\Phi\left(z-\frac{\sigma_0(\theta)}{2}\right) \right)
\right\}\right\}
= \phi\left(z + \frac{\sigma_0(\theta)}{2}\right).
\end{equation*}
The derivative of $H_k(z)\phi(z)$ is $-H_{k+1}(z)\phi(z)$.
Recall also $\sigma_0(\theta) = O(\sqrt{\theta})$.
 \hfill////
  
  \subsection{Implied volatility expansion}
  Here we prove an expansion formula for the Black-Scholes implied
  volatility. 
\\

\noindent
    {\it Proof  of Theorem~\ref{thm:ivex}: }
Step 1).
Fix $z \in \mathbb{R}$.  Note that
\begin{equation} \label{bsput}
P_{\theta}(\sigma):= \frac{ p_{\rm{BS}}(Fe^{\sqrt{\theta}z},\theta,\sigma) }{Fe^{-r\theta}\sqrt{\theta}}= 
\frac{1}{\sqrt{\theta}}\left(\Phi\left(\frac{z}{\sigma} + \frac{\sigma \sqrt{\theta}}{2}\right)e^{\sqrt{\theta}z}-\Phi\left(\frac{z}{\sigma}-\frac{\sigma\sqrt{\theta}}{2}\right) \right)
\end{equation}
and that
\begin{equation*}
 P_\theta : [0,\infty] \to \left[
\frac{ (e^{\sqrt{\theta}z}-1)_+}{\sqrt{\theta}}, \frac{ e^{\sqrt{\theta}z}}{\sqrt{\theta}} 
\right]
\end{equation*}
is a strictly increasing function.
From (\ref{bsput}) and Proposition~\ref{thm:prex}, we have
  \begin{equation*}
  \begin{split}
  \frac{ p(Fe^{\sqrt{\theta}z},\theta) }{Fe^{-r\theta}\sqrt{\theta}}
  =   &  P_\theta(\kappa_2)
+ \kappa_2 \kappa_3\phi \left(\frac{z}{\kappa_2} + \frac{\kappa_2\sqrt{\theta}}{2}\right)
H_1\left(\frac{z}{\kappa_2} + \frac{\kappa_2\sqrt{\theta}}{2}\right)e^{\sqrt{\theta}z} \theta^{H} 
\\
& + \kappa_2 \phi\left(\frac{z}{\kappa_2}\right)\left(\kappa_4H_2\left(\frac{z}{\kappa_2}\right) + \frac{\kappa_3^2}{2}H_4\left(\frac{z}{\kappa_2}\right) \right)\theta^{2H}
  + o(\theta^{2H}) \\
 = &  P_\theta(\kappa_2) + O(\theta^H).
  \end{split}
  \end{equation*}
Therefore
\begin{equation*}
 \sigma_{\rm{BS}}(\sqrt{\theta}z,\theta) =
P_\theta^{-1}(P_{\theta}(\kappa_2) + O(\theta^H)).
\end{equation*}
By (\ref{smooth}), $\kappa_2$ is bounded in $\theta$, say, by $L > 0$.
The function $P_\theta$ converges as $\theta \to 0$ to 
\begin{equation*}
 P_0(\sigma) := z \Phi\left(\frac{z}{\sigma}\right) + \sigma \phi \left(\frac{z}{\sigma}\right)
\end{equation*}
pointwise, and by Dini's theorem, this convergence is uniform on $[0,L]$.
Since the limit function $P_0$ is strictly increasing,
the inverse functions $P^{-1}_\theta$ converges to $P_0^{-1}$.
Again by Dini's theorem, this convergence is uniform and in particular,
$P_\theta^{-1}$ are equicontinuous.
Thus we conclude $\sigma_{\rm{BS}}(\sqrt{\theta}z,\theta)  - \kappa_2 \to 0$ as $\theta \to 0$. Then, 
write  $\sigma_{\rm{BS}}(\sqrt{\theta}z,\theta) = \kappa_2 + \beta(\theta)$
and substitute this to the equation 
$P_\theta(\sigma_{\rm{BS}}(\sqrt{\theta}z,\theta)) = P_\theta(\kappa_2) + O(\theta^H)$. The Taylor expansion gives
$\beta(\theta) = O(\theta^H)$.
\\

\noindent Step 2).
From (\ref{bsput}) we have
  \begin{equation*}
 P_\theta(\sigma)  = 
    \sigma F_1\left(\frac{z}{\sigma}\right) +
    \frac{\sigma^2 \sqrt{\theta}}{2}
    F_2\left(\frac{z}{\sigma}\right) +
     \frac{ \sigma^3 \theta}{6}
    F_3\left(\frac{z}{\sigma}\right) +
    o(\theta),
   \end{equation*}
   where
   \begin{equation*}
   F_1(x) = x\Phi(x) + \phi(x), \ \  F_2(x) = x^2 \Phi(x) + x\phi(x), \ \ 
      F_3(x) = x^3 \Phi(x) + \left(x^2-\frac{1}{4}\right)\phi(x).
     \end{equation*}
    Using that
    \begin{equation*}
    \partial_\sigma \left\{
\sigma F_1 \left(\frac{z}{\sigma}\right) 
    \right\} 
    =  \phi\left(\frac{z}{\sigma}\right),
    \end{equation*}
    we have
 \begin{equation*}
  \begin{split}
&  \kappa_2 F_1\left(\frac{z}{\kappa_2}\right) +
    \frac{\kappa_2^2 \sqrt{\theta}}{2}  F_2\left(\frac{z}{\kappa_2}\right) + 
  \kappa_2\phi\left(\frac{z}{\kappa_2}\right)
  \kappa_3H_1\left(\frac{z}{\kappa_2}\right)e^{\sqrt{\theta}z}
  \theta^{H}
  \\& = \sigma_{\rm{BS}}(\sqrt{\theta}z,\theta)
  F_1\left(\frac{z}{\sigma_{\rm{BS}}(\sqrt{\theta}z,\theta)}\right) +
    \frac{\sigma_{\rm{BS}}(\sqrt{\theta}z,\theta)^2 \sqrt{\theta}}{2}
    F_2\left(\frac{z}{\sigma_{\rm{BS}}(\sqrt{\theta}z,\theta)}\right)
    + O(\theta^{2H})
    \\ & =
    \kappa_2 F_1\left(\frac{z}{\kappa_2}\right) +
    \frac{\kappa_2^2 \sqrt{\theta}}{2}
    F_2\left(\frac{z}{\kappa_2}\right) +
    \phi\left(\frac{z}{\kappa_2}\right)(\sigma_{\rm{BS}}(\sqrt{\theta}z,\theta)-\kappa_2)
    + O(\theta^{2H}),
  \end{split}
  \end{equation*}
  from which we conclude
  $\sigma_{\rm{BS}}(\sqrt{\theta}z,\theta) =
  \kappa_2 + \kappa_3 z e^{\sqrt{\theta}z}\theta^H + O(\theta^{2H})$.\\

 \noindent
Step 3).  Using that
      \begin{equation*}
\partial_\sigma^2 \left\{
\sigma F_1 \left(\frac{z}{\sigma}\right) 
    \right\} 
    =   \frac{z^2}{\sigma^3} \phi\left(\frac{z}{\sigma}\right), \ \ 
    \partial_\sigma \left\{
\sigma^2 F_2 \left(\frac{z}{\sigma}\right) 
    \right\} 
    =  z \phi\left(\frac{z}{\sigma}\right),
    \end{equation*}
    we obtain
      \begin{equation*}
      \begin{split}
 \kappa_2\phi &\left(\frac{z}{\kappa_2} + \frac{\kappa_2 \sqrt{\theta}}{2}\right)  \left(
  \kappa_3H_1\left(\frac{z}{\kappa_2} + \frac{\kappa_2 \sqrt{\theta}}{2} \right)e^{\sqrt{\theta}z} \theta^{H}
 + \left(\kappa_4H_2\left(\frac{z}{\kappa_2}\right) +
      \frac{\kappa_3^2}{2}H_4\left(\frac{z}{\kappa_2}\right)
      \right)\theta^{2H}\right)  \\
 = &  \frac{ p(Fe^{\sqrt{\theta}z},\theta) }{Fe^{-r\theta}\sqrt{\theta}}
 -  P_\theta(\kappa_2) + o(\theta^{2H}) \\
=& 
P_\theta(\sigma_{\rm{BS}}(\sqrt{\theta}z,z))
 -  P_\theta(\kappa_2) + o(\theta^{2H}) \\
=& 
\partial_\sigma \left\{
\sigma F_1 \left(\frac{z}{\sigma}\right) 
    \right\}\big|_{\sigma = \kappa_2}  (\sigma_{\rm{BS}}(\sqrt{\theta}z,\theta)-\kappa_2) + \frac{1}{2}\partial_\sigma^2 \left\{
\sigma F_1 \left(\frac{z}{\sigma}\right) 
    \right\}\big|_{\sigma = \kappa_2} (\sigma_{\rm{BS}}(\sqrt{\theta}z,\theta)-\kappa_2)^2\\
 & +  \frac{\sqrt{\theta}}{2}  \partial_\sigma \left\{
\sigma^2 F_2 \left(\frac{z}{\sigma}\right) 
    \right\} \bigg|_{\sigma=\kappa_2} (\sigma_{\rm{BS}}(\sqrt{\theta}z,\theta)-\kappa_2) + o(\theta^{2H})\\
      = &
          \phi\left(\frac{z}{\kappa_2}\right)(\sigma_{\rm{BS}}(\sqrt{\theta}z,\theta)-\kappa_2)
    	  + \frac{\sqrt{\theta}}{2}
	  z
      \phi\left(\frac{z}{\kappa_2}\right)(\sigma_{\rm{BS}}(\sqrt{\theta}z,\theta)-\kappa_2)
      \\
      & + \frac{z^2}{2\kappa_2^3}\phi\left(\frac{z}{\kappa_2}\right)(\sigma_{\rm{BS}}(\sqrt{\theta}z,\theta)-\kappa_2)^2
  + o(\theta^{2H})
  \end{split}
  \end{equation*}
from Theorem~\ref{thm:prex} and Step 2.
The left hand side is further expanded as
\begin{equation*}
\begin{split}
\kappa_2 \phi\left(\frac{z}{\kappa_2}\right) \Biggl\{
 \kappa_3 H_1\left(\frac{z}{\kappa_2}\right)e^{\sqrt{\theta}z} \theta^H - & \kappa_3 H_2\left(\frac{z}{\kappa_2}\right) \frac{\kappa_2 }{2} \theta^{H+1/2} \\
& +  \left(\kappa_4H_2\left(\frac{z}{\kappa_2}\right) +
      \frac{\kappa_3^2}{2}H_4\left(\frac{z}{\kappa_2}\right)
      \right)\theta^{2H}
 \Biggr\}  + o(\theta^{2H}).
\end{split}
\end{equation*}
Denote $\gamma(\theta) = \sigma_{\rm{BS}}(\sqrt{\theta}z,\theta) -
  \kappa_2 - \kappa_3 z e^{\sqrt{\theta}z}\theta^H$ and substitute this to obtain
\begin{equation*}
\begin{split}
 \gamma(\theta) = &
- \kappa_3 H_2\left(\frac{z}{\kappa_2}\right) \frac{\kappa_2^2 }{2} \theta^{H+1/2} +  \kappa_2 \left(\kappa_4H_2\left(\frac{z}{\kappa_2}\right) +
      \frac{\kappa_3^2}{2}H_4\left(\frac{z}{\kappa_2}\right)
      \right)\theta^{2H}
 \\ &-\frac{\kappa_3 }{2} z^2 \theta^{H+1/2}-
\frac{\kappa_3^2}{2\kappa_2^3} z^4\theta^{2H} + o(\theta^{2H}) \\
= & \left(\frac{\kappa_2^2}{2}-z^2\right)\kappa_3\theta^{H+1/2}
+\kappa_2
 \left((\kappa_4-3\kappa_3^2)\frac{z^2}{\kappa_2^2} + \frac{3}{2}\kappa_3^2-\kappa_4\right) \theta^{2H} + o(\theta^{2H}),
\end{split}
\end{equation*}
  from which we conclude the result. \hfill////

  \subsection{Asymptotics for at-the-money skew and curvature}
  Here we prove Theorem~\ref{thm:asbe}.
\\

\noindent
      {\it Proof of Theorem~\ref{thm:asbe}: }
  It is known (see e.g., Fukasawa~\cite{MF}) that
  \begin{equation}\label{ders}
  \begin{split}
  &  \partial_k \sigma_{\rm{BS}}(k,\theta)
  = \frac{Q(k \geq \sigma_0(\theta)X_\theta ) -
  \Phi(f_2(k,\theta))}{\sqrt{\theta}\phi(f_2(k,\theta))}, \\
  &   \partial_k^2 \sigma_{\rm{BS}}(k,\theta)
  = \frac{p_\theta(k /\sigma_0(\theta))}{ \sigma_0(\theta)\sqrt{\theta} \phi(f_2(k,\theta))} -
  \sigma_{\rm{BS}}(k,\theta) \partial_kf_1(k,\theta)
     \partial_kf_2 (k,\theta),
  \end{split}
  \end{equation}
  where
  \begin{equation*}
  f_1(k,\theta) = \frac{k}{\sqrt{\theta}\sigma_{\rm{BS}}(k,\theta)}
  - \frac{\sqrt{\theta}\sigma_{\rm{BS}}(k,\theta)}{2}, \ \
  f_2(k,\theta) = \frac{k}{\sqrt{\theta}\sigma_{\rm{BS}}(k,\theta)}
  + \frac{\sqrt{\theta}\sigma_{\rm{BS}}(k,\theta)}{2}.
  \end{equation*}
Since the condition of Theorem~\ref{thm:prex} is met, 
we have
  \begin{equation*}
  Q(0 \geq X_\theta)
  = \Phi\left(\frac{\sigma_0(\theta)}{2}\right) + \kappa_3(\theta)\phi\left(\frac{\sigma_0(\theta)}{2}\right) \theta^H + o(\theta^{2H}).
  \end{equation*}
  On the other hand, by Theorem~\ref{thm:ivex},
  \begin{equation*}
  f_2(0,\theta) = \frac{\sqrt{\theta}}{2} \kappa_2(\theta) +
  O(\theta^{2H+1/2})
  \end{equation*}
  and so,
  \begin{equation*}
  \begin{split}
&  \Phi(f_2(0, \theta))
  = \Phi\left(\frac{\sigma_0(\theta)}{2}\right) +
  O(\theta^{2H+1/2}),
  \\
&  \phi(f_2(0,\theta)) = \phi(0) - \phi(0)\frac{\theta}{8}\kappa_2(\theta)^2 +
  O(\theta^{2H+1}).
 \end{split} \end{equation*}
Then, it follows from (\ref{ders}) that
     \begin{equation} \label{skew}
      \partial_k \sigma_{\rm{BS}}(0,\theta)
      = \kappa_3(\theta)\theta^{H-1/2} + o(\theta^{2H-1/2}).
 \end{equation} 
 Further, under the condition, we have
 \begin{equation*}
 p_\theta(0)
 = \phi\left(\frac{\sigma_0(\theta)}{2}\right)\left\{1 -\frac{ \kappa_3(\theta)}{2} \sigma_0(\theta)\theta^H
  + \left(3\kappa_4(\theta) - 15 \frac{\kappa_3(\theta)^2}{2}\right)\theta^{2H}\right\}
  + o(\theta^{2H}).
  \end{equation*}
    On the other hand, by Theorem~\ref{thm:ivex} and (\ref{skew}),
    \begin{equation*}
    \begin{split}
    \sigma_{\rm{BS}}(0,\theta) &
\partial_kf_1(0,\theta) \partial_kf_2(0,\theta) \\ = &
    \frac{1}{\sigma_{\rm{BS}}(0,\theta) \theta} + O(\theta^{2H})
   \\  = & \frac{1}{\kappa_2(\theta) \theta}
    \left(1 -\frac{1}{2} \kappa_2(\theta) \kappa_3(\theta)\theta^{H+1/2}
  -  \left(\frac{3}{2}\kappa_3(\theta)^2-\kappa_4(\theta)\right)\theta^{2H}\right)
    + o(\theta^{2H-1}).
    \end{split}
    \end{equation*}
    Then, it follows from (\ref{ders}) that
    \begin{equation*}
    \partial_k^2 \sigma_{\rm{BS}}(0,\theta) =
    \frac{2\kappa_4(\theta) - 6 \kappa_3(\theta)^2}{\kappa_2(\theta)} \theta^{2H-1}
    + o(\theta^{2H-1}),
    \end{equation*}
    which completes the
    proof. \hfill////\\

\section{Regular stochastic volatility models}
Here we briefly discuss that regular stochastic
volatility models satisfy all the assumptions in Section~2.1 with $H = 1/2$.
Consider the volatility process  $v_t = v(X_t)$, where
 $X$ is a Markov process
satisfying a stochastic differential equation
\begin{equation*}
 \mathrm{d}X_t = b(X_t) \mathrm{d}t + c(X_t)\mathrm{d}W_t
\end{equation*}
and $v$ is a smooth positive function defined
on the state space of $X$.
Let $\rho \in (-1,1)$ be a constant and $\{\mathcal{G}_t\}$ be the augmented
filtration generated by $W$.
We assume (\ref{smooth}), which is satisfied in the usual cases including the
log-normal SABR and Heston models. 
Denote by $L$ the generator of $X$.
Put $f = \sqrt{v}$, $g = f^\prime c$ and $h = v^\prime c$. 
Then, by It\^o's formula,
we have
\begin{equation*}
 \begin{split}
& M_\theta = f(X_0)B_\theta + \int_0^\theta
  \int_0^t g(X_s)\mathrm{d}W_s\mathrm{d}B_t + 
    \int_0^\theta\int_0^t Lf(X_s)\mathrm{d}s\mathrm{d}B_t,\\
  & \langle M \rangle_\theta
  = v(X_0) \theta +
  \int_0^\theta \int_0^t h(X_s)\mathrm{d}W_s\mathrm{d}t + 
    \int_0^\theta\int_0^t Lv(X_s)\mathrm{d}s\mathrm{d}t.
 \end{split}
\end{equation*}
Let $\bar{B}^\theta_t = \theta^{-1/2}B_{\theta t}$, $\bar{W}^\theta_t =
\theta^{-1/2}W_{\theta t}$ and $X^\theta_t = X_{\theta t}$.
Then
\begin{equation*}
 \begin{split}
& \frac{M_\theta}{\sqrt{\theta}}
  = f(X_0)\bar{B}^\theta_1 + \sqrt{\theta} \int_0^1
  \int_0^ug(X^\theta_v)\mathrm{d}\bar{W}^\theta_v
  \mathrm{d}\bar{B}^\theta_u +
   \theta \int_0^1
  \int_0^uLf(X^\theta_v)\mathrm{d}v
  \mathrm{d}\bar{B}^\theta_u,\\
  &
  \frac{\langle M \rangle_\theta}{\theta}
  = v(X_0) + \sqrt{\theta} \int_0^1
  \int_0^uh(X^\theta_v)\mathrm{d}\bar{W}^\theta_v
  \mathrm{d}u +
   \theta \int_0^1
  \int_0^uLv(X^\theta_v)\mathrm{d}v
  \mathrm{d}u.
 \end{split}
\end{equation*}
It  would follow that
\begin{equation*}
 \frac{\sigma_0(\theta)^2}{\theta}
  = \frac{E[\langle M \rangle_\theta]}{\theta}
  = v(X_0) + \frac{1}{2}Lv(X_0) \theta + O(\theta^{3/2}),
\end{equation*}
and so
\begin{equation*}
 \frac{\sigma_0(\theta)}{\sqrt{\theta}}
  = f(X_0) + \frac{1}{4} \frac{Lv(X_0)}{f(X_0)} \theta + O(\theta^{3/2})
\end{equation*}
under a mild regularity condition.
Then, we have (\ref{stex}) with $H = 1/2$,
$M^{(0)}_\theta = \bar{B}^\theta_1$ and
\begin{equation*}
 \begin{split}
& M^{(1)}_\theta = \frac{g(X_0)}{f(X_0)} \int_0^1 \bar{W}^\theta_u
  \mathrm{d}\bar{B}^\theta_u, \\
&  M^{(2)}_\theta =
  -\frac{Lv(X_0)}{4v(X_0)}\bar{B}^\theta_1
   + \frac{g^{\prime}(X_0)c(X_0)}{f(X_0)}\int_0^1 \int_0^u 
   \bar{W}^\theta_v
   \mathrm{d}\bar{W}^\theta_v
  \mathrm{d}\bar{B}^\theta_u +
  \frac{Lf(X_0)}{f(X_0)}\int_0^1 u
  \mathrm{d}\bar{B}^\theta_u,
  \\
    & M^{(3)}_\theta = 2\frac{g(X_0)}{f(X_0)} \int_0^1
  \bar{W}^\theta_u \mathrm{d}u
 \end{split}
\end{equation*}
again under a mild regularity condition.
By Nualart et al.~\cite{NUZ} or Appendix A below,
\begin{equation*}
 E[M^{(1)}_\theta|M^{(0)}_\theta = x]
= \frac{g(X_0)}{f(X_0)} \frac{\rho}{2}H_2(x), \ \ 
 E[M^{(3)}_\theta|M^{(0)}_\theta = x]
= \frac{g(X_0)}{f(X_0)} \rho H_1(x)
\end{equation*}
and so,
\begin{equation*}
\begin{split}
 a^{(1)}_\theta & \left(x + \frac{\sigma_0(\theta)}{2}\right) -
\frac{\sigma_0(\theta)}{2} a^{(3)}_\theta
\left(x + \frac{\sigma_0(\theta)}{2}\right)
\\ & = - \kappa_3
\left(H_3 \left(x + \frac{\sigma_0(\theta)}{2}\right)
- \sigma_0(\theta)H_2 \left(x + \frac{\sigma_0(\theta)}{2}\right)
\right)\phi\left(x + \frac{\sigma_0(\theta)}{2}\right)
\end{split}
\end{equation*}
with
\begin{equation*}
 \kappa_3  =\frac{\rho}{2}  \frac{g(X_0)}{f(X_0)}.
\end{equation*}
Further,
\begin{equation*}
\begin{split}
 E[M^{(2)}_\theta|M^{(0)}_\theta = x]
&= -\frac{Lv}{4v}(X_0)x + \frac{g^\prime c}{f}(X_0) \frac{\rho^2}{6}H_3(x)
+ \frac{Lf}{2f}(X_0) x
\\ & = - \frac{g^2}{4f^2}(X_0)x + \frac{g^\prime c}{f}(X_0) \frac{\rho^2}{6}H_3(x) 
\end{split}
\end{equation*}
and
\begin{equation*}
  E[(M^{(1)}_\theta)^2|M^{(0)}_\theta = x] =
\frac{g^2}{f^2}(X_0)\left(\rho^2 \left(\frac{1}{2} + H_2(x)+ \frac{1}{4}H_4(x) \right) + (1-\rho^2)\left(\frac{1}{2} + \frac{1}{3}H_2(x)\right)\right).
\end{equation*}
Therefore,
\begin{equation*}
 a^{(2)}_\theta(x) - \frac{1}{2}c_\theta(x) = 
-\kappa_4 H_4(x)\phi(x) - \frac{\kappa_3 ^2}{2}H_6(x)\phi(x)
\end{equation*}
with
\begin{equation*}
 \kappa_4  =  \frac{g^\prime c}{f}(X_0) \frac{\rho^2}{6}
+ \frac{g^2}{f^2}(X_0)\frac{1+ 2\rho^2}{6}.
\end{equation*}
Thus we have observed that (\ref{qth2}) has the form of (\ref{thm32form}) and so, all the theorems in Section~2.2 are applied.
In particular, Theorem~\ref{thm:ivex} proves the Medvedev-Scaillet formula (Proposition 1 of \cite{MS0} that was obtained by a formal computation when $f$ is the identity function).

    \section{The rough Bergomi model}
    Here we show that the rough Bergomi model proposed by \cite{BFG}
 fits into the framework and compute the expansion terms.
 Let $\rho_t = \rho \in (-1,1)$ be a constant and 
  \begin{equation*}
  v_t = v_0(t) \exp\left\{
\eta\sqrt{2H} \int_0^t (t-s)^{H-1/2} \mathrm{d}W_s -\frac{\eta^2}{2} t^{2H}
  \right\}.
  \end{equation*}
The deterministic function $v_0(t) =E[v_t]$ is assumed to be continuous.
  \begin{thm} \label{thm:rbergomi}
  We have (\ref{denex}) for $q_\theta$ given by 
(\ref{thm32form}) with
  \begin{equation*}
  \begin{split}
&  \kappa_3(\theta)
  =\rho \eta \sqrt{\frac{H}{2}}
  \frac{1}{\theta^H\sigma_0(\theta)^3}
  \int_0^{\theta}\int_0^t
  (t-s)^{H-1/2}\sqrt{v_0(s)}\mathrm{d}s v_0(t)\mathrm{d}t,
  \\
  & \kappa_4(\theta) =   \frac{(1+2\rho^2)\eta^2 H}{(2H+1)^2(2H+2)}
  +  \frac{\rho^2\eta^2 H \beta(H+3/2,H+3/2)}{2(H+1/2)^2},
  \end{split}
  \end{equation*}
where $\beta$ is the beta function.
  \end{thm}
{\it Proof: }
 Since $v_t$ is log-normally distributed, (\ref{smooth}) holds by Jensen's inequality. 
The conditions (\ref{moments}) and (\ref{stex}) follow from
Lemma~\ref{lemrB} below. The functions $a^{(i)}_\theta$ and $c_\theta$
are computed in Lemmas~\ref{lemrB2}, \ref{lemrB3}, \ref{lemrB4} and \ref{lemrB5} below. The function $b_\theta$ is obtained as the 
the derivative of $a^{(1)}_\theta$.  They are apparently rapidly decreasing smooth functions.
Then, by Theorem~\ref{thm:denex}, it suffices to show that $q_\theta$ defined by
(\ref{qth}) has the form (\ref{thm32form}) up to $o(\theta^{2H})$ with 
$\kappa_3(\theta)$ and $\kappa_4(\theta)$ specified above.
\hfill////\\

 Theorems~\ref{thm:prex}, \ref{thm:ivex} and  \ref{thm:asbe}
 are therefore valid here as well.
When $H < 1/2$ and
  the forward variance curve is flat (i.e., $v_0$ is constant),
Theorem~\ref{thm:ivex} gives a similar formula to 
the Bergomi-Guyon expansion formally derived in \cite{BFG}\footnote{ 
 Note however that there is a typo in the second order term in \cite{BFG}.}.
In fact, the expansion of $O(\eta)$ given in \cite{BFG} coincides with
our expansion of $O(\theta^H)$ when $v_0$ is constant.
When $v_0$ is not constant, or when looking at the second-order terms,
the formulas are not the same; this is not surprising 
because the asymptotics are 
$\eta \to 0$ in \cite{BFG}  while $\theta \to 0$ here. 
  Further, when $v_0$ is constant,  the same formula of $O(\theta^H)$ can 
be  obtained by expanding the rate function of the large deviation result of  \cite{FZ} as in \cite{Friz}. To be more precise, note that by Theorem~\ref{thm:rbergomi},
\begin{equation*}
 \kappa_3(\theta) = \rho \frac{\eta\sqrt{2H}}{2(H+1/2)(H+3/2)}
\end{equation*}
when $v_0$ is constant and Theorem~\ref{thm:ivex} implies
\begin{equation*}
 \frac{\sigma_{\mathrm{BS}}(\sqrt{\theta}z,\theta) - \sigma_{\mathrm{BS}}(\sqrt{\theta}\zeta,\theta) }{\sqrt{\theta}z - \sqrt{\theta}\zeta} = \kappa_3(\theta)\theta^{H-1/2} + O(\theta^{2H-1/2})
\end{equation*}
for $z \neq \zeta$. A weaker assertion, where 
$O(\theta^{2H-1/2})$ is replaced with $o(\theta^{H-1/2})$,
 was already shown in \cite{F2017} by a different method.
What is shown in \cite{Friz} via an expansion of the rate function 
 is that this formula up to  $o(\theta^{H-1/2})$ remains valid even 
if $\sqrt{\theta} z$ and 
$\sqrt{\theta} \zeta$ are replaced with
$\theta^\beta z$ and 
$\theta^\beta \zeta$ respectively for $\beta \in (1/2-H,1/2)$.\\

\begin{figure}
\includegraphics[bb=0 0 504 504, width=5.8cm]{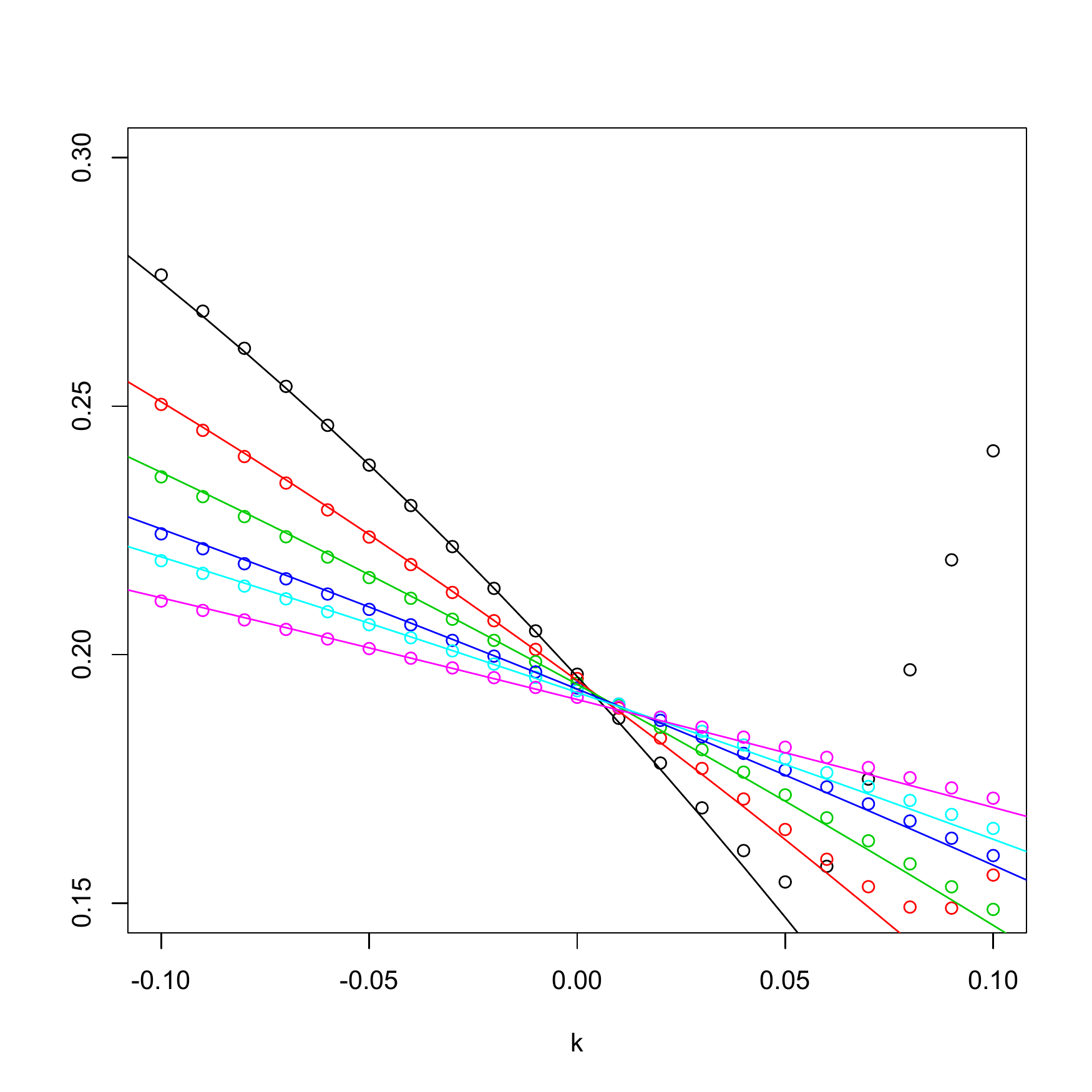}
\includegraphics[bb=0 0 504 504, width=5.8cm]{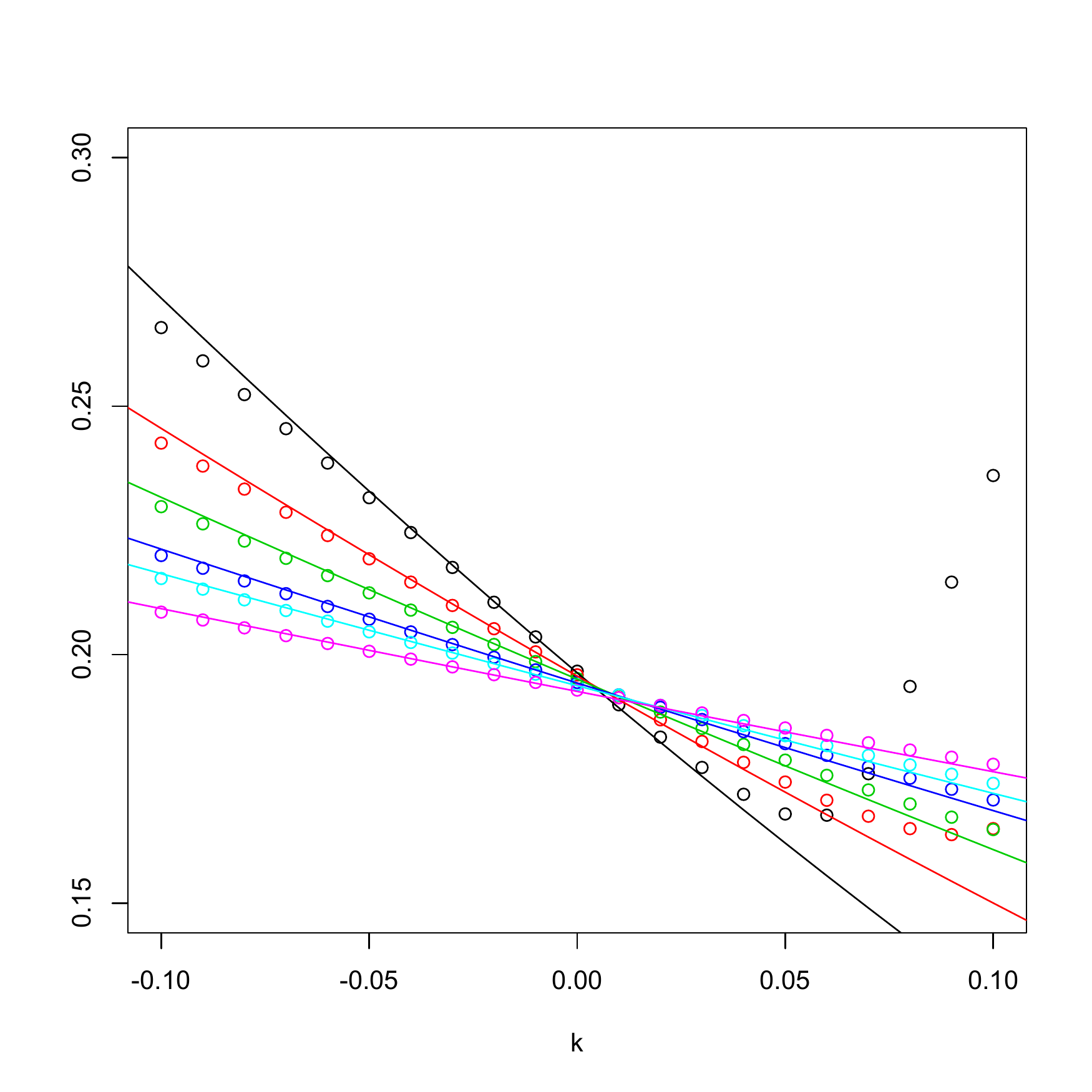}
\caption{The Black-Scholes implied volatility under the rough Bergomi model with $v_0 \equiv .04$ and $(H,\rho,\eta) = (.07,-.9,.9)$ (left) or 
$(H,\rho,\eta) = (.07,-.7,.9)$ (right).  }\label{t1}
\end{figure}

\begin{figure}
\includegraphics[bb=0 0 504 504, width=5.8cm]{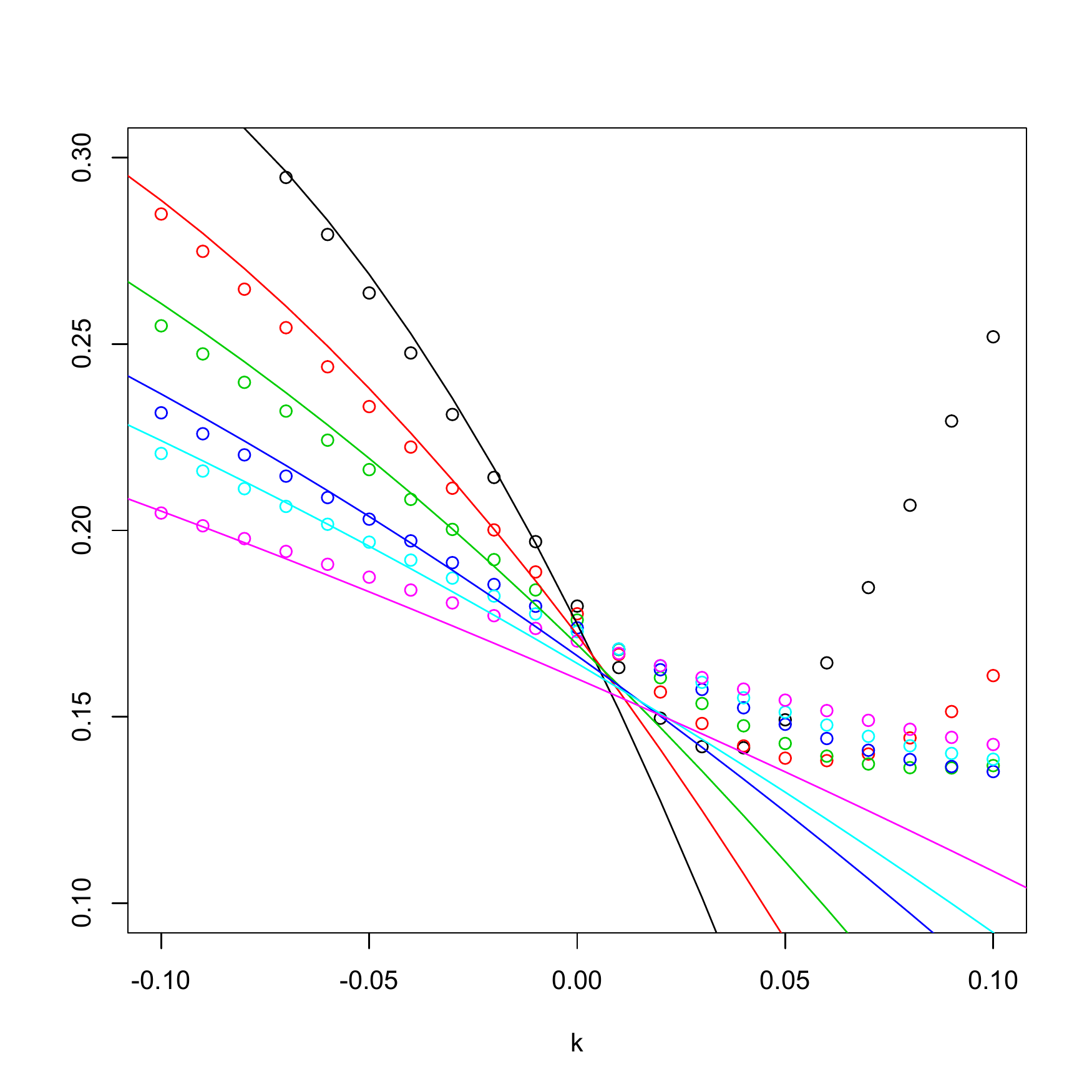}
\includegraphics[bb=0 0 504 504, width=5.8cm]{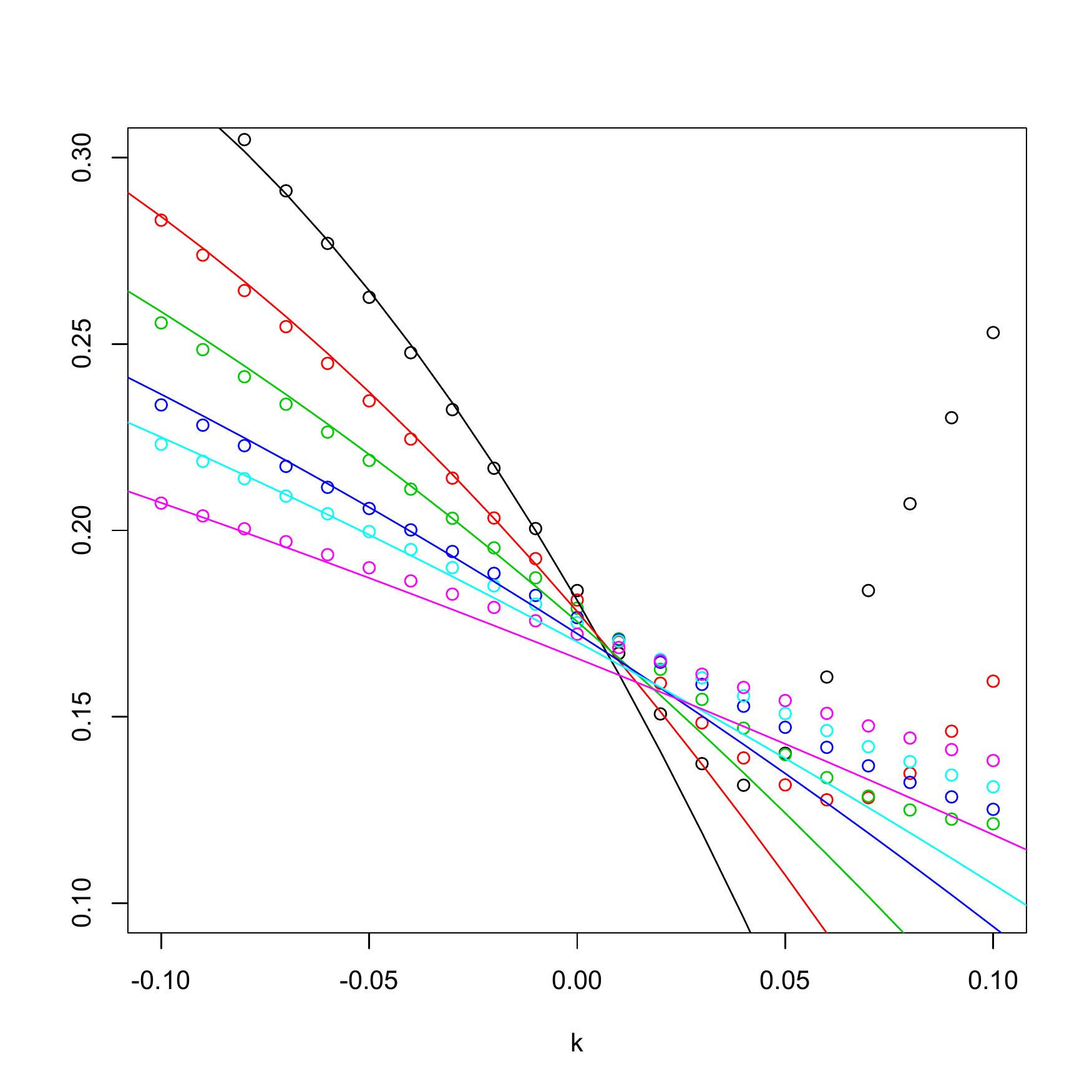}
\caption{The Black-Scholes implied volatility under the rough Bergomi model with $v_0 \equiv .04$ and $(H,\rho,\eta) = (.05,-.9,2.3)$ (left) or 
$(H,\rho,\eta) = (.07,-.9,1.9)$ (right).  }\label{t2}
\end{figure}

How small $\theta$ has to be for reasonable accuracy of our asymptotic formulas should be examined via numerical experiments. Our extensive experiments suggest
 $\eta \theta^H < 1$ as a rough criterion\footnote{Note that $\eta \theta^H$ is the standard deviation of log-spot-variance.}.  Here we present only a few examples of the volatility surfaces. In Figures~\ref{t1} and \ref{t2}, the points are by the Monte Carlo and the curves are by the asymptotic formula given in Theorems~\ref{thm:ivex}  and \ref{thm:rbergomi}.
The different colors are for different time-to-maturities;
black for $\theta = .02$, red for $\theta = .05$, green for $\theta=.1$,
blue for $\theta = .2$, cyan for $\theta = .3$ and magenta for $\theta = .6$.
Note that the sets of  parameters  in Figure~\ref{t2}
 are those calibrated from option data by Bayer et al~\cite{BFG}.
\\

In order to prove Lemmas below, we need some preparation. 
Let $H_k$, $k=0,1,\dots$ be the Hermite polynomials as before:
  \begin{equation*}
  H_k(x) = (-1)^k e^{x^2/2}\frac{\mathrm{d}^k}{\mathrm{d}x^k}
  e^{-x^2/2}
  \end{equation*}
  and $H_k(x,a) = a^{k/2}H_k(x/\sqrt{a})$ for $a > 0$.
  As is well-known, we have
  \begin{equation*}
  \exp\left\{ux - \frac{au^2}{2}\right\} = \sum_{k=0}^\infty
  H_k(x,a)\frac{u^k}{k!}
  \end{equation*}
  and for any continuous local martingale $M$ and $n \in \mathbb{N}$,
  \begin{equation}\label{hel}
  \mathrm{d} L^{(n)}_t = n L^{(n-1)}_t \mathrm{d}M_t,
  \end{equation}
  where $L^{(k)} = H_k(M,\langle M \rangle)$ for $k \in \mathbb{N}$.
  See, e.g., Revuz and Yor~\cite{RY}.

  Define $\hat{W}$, $\hat{W}^\prime$, $\hat{B}$ by
  \begin{equation*}
  \hat{W}_t =\frac{1}{\sigma_0(\theta)}
  \int_0^{\tau^{-1}(t)}\sqrt{v_0(s)}
  \mathrm{d}W_s, \ \ 
  \hat{W}^\prime_t =\frac{1}{\sigma_0(\theta)}
  \int_0^{\tau^{-1}(t)}\sqrt{v_0(s)}
  \mathrm{d}W^\prime_s
  \end{equation*}
  and $\hat{B} = \rho \hat{W} + \sqrt{1-\rho^2}\hat{W}^\prime$,
  where
  \begin{equation*}
  \tau(s) = \frac{1}{\sigma_0(\theta)^2}\int_0^s v_0(t)\mathrm{d}t.
  \end{equation*}
  Then, $(\hat{W},\hat{W}^\prime)$ is a 2-dimensional Brownian motion under $E$
  and
for any square-integrable function $f$,
  \begin{equation*}
  \int_0^a f(s)\mathrm{d}W_s
  =
  \sigma_0(\theta)\int_0^{\tau(a)}\frac{f(\tau^{-1}(t))}{\sqrt{v_0(\tau^{-1}(t))}}\mathrm{d}\hat{W}_t.
  \end{equation*}
  Therefore,
  \begin{equation*}
  M_\theta =
  \sigma_0(\theta)\int_0^1
  \exp\left\{
\theta^HF^t_t
  - \frac{\eta^2}{4}|\tau^{-1}(t)|^{2H}
  \right\}\mathrm{d}\hat{B}_t
  \end{equation*}
  where
  \begin{equation*}
  F^t_u =
  \eta\sqrt{\frac{H}{2}} \frac{\sigma_0(\theta)}{\theta^H}\int_0^u
  \frac{(\tau^{-1}(t)-\tau^{-1}(s))^{H-1/2}}{\sqrt{v_0(\tau^{-1}(s))}}\mathrm{d}\hat{W}_s,
  \ \ u \in [0,t].
  \end{equation*}
  Let
  \begin{equation*}
  G^{(k)}_t = H_k(F^t_t,\langle F^t \rangle_t).
  \end{equation*}
 Then, we have
  \begin{equation*}
  \begin{split}
  M_\theta &=
\sigma_0(\theta)\int_0^1
  \exp\left\{-\frac{\eta^2}{8}|\tau^{-1}(t)|^{2H}\right\}
  \exp\left\{\theta^H F^t_t - \frac{\theta^{2H}}{2}\langle F^t \rangle_t
  \right\} \mathrm{d}\hat{B}_t\\
& =  \sigma_0(\theta)\int_0^1
  \exp\left\{-\frac{\eta^2}{8}|\tau^{-1}(t)|^{2H}\right\}
  \sum_{k=0}^{\infty}G^{(k)}_t\frac{\theta^{Hk}}{k!}  \mathrm{d}\hat{B}_t.
 \end{split} \end{equation*}
   \begin{lem} \label{lemrB}
  We have (\ref{stex}) with
  \begin{equation*}
  \begin{split}
  & M^{(0)}_\theta = \hat{B}_1, \\
    & M^{(1)}_\theta = 
  \int_0^1  h_\theta(t)
  G^{(1)}_t \mathrm{d}\hat{B}_t, \\
 & M^{(2)}_\theta =
  \int_0^1  \left\{
\frac{h_\theta(t)-1}{\theta^{2H}} 
  + h_\theta(t) \frac{
  G^{(2)}_t}{2}\right\} \mathrm{d}\hat{B}_t, \\
   & M^{(3)}_\theta = 2
   \int_0^1 F^t_t
   \mathrm{d}t, 
  \end{split}
  \end{equation*}
  where
  \begin{equation*}
  h_\theta (t)
  =  \exp\left\{-\frac{\eta^2}{8}|\tau^{-1}(t)|^{2H}\right\}.
  \end{equation*}
  \end{lem}
  {\it Proof: }
  For $M^{(i)}_\theta$, $i=0,1,2$,
  it suffices to show
  \begin{equation*}
  \left\|
  \int_0^1 h_\theta(t)
  \sum_{k=J}^\infty G^{(k)}_t \frac{\theta^{Hk}}{k!} \mathrm{d}\hat{B}_t
  \right\|_2 = O(\theta^{HJ})
  \end{equation*}
  for any $J \geq 3$. The proof for $M^{(3)}_\theta$ is similar and so omitted.
  It suffices to show
  \begin{equation*}
  E\left[\int_0^1 
  \left| \sum_{k=J}^\infty G^{(k)}_t \frac{\theta^{Hk}}{k!} \right|^2
  \mathrm{d}t \right] = O(\theta^{2HJ}).
  \end{equation*}
  By the Cauchy-Schwarz inequality,  the left hand side is dominated by
  \begin{equation*}
  \sum_{k=J}^\infty \theta^{Hk}
  \sum_{k=J}^\infty \frac{\theta^{Hk}}{(k!)^2}\int_0^1
  E[|G^{(k)}_t|^2]\mathrm{d}t
  \end{equation*}
  Let
  \begin{equation*}
  G^{(k)}_{t,s} = H_k(F^t_s,\langle F^t \rangle_s), \ \ s \in [0,t].
  \end{equation*}
  Then, by (\ref{hel}),
  \begin{equation*}
  \begin{split}
  E[|G^{(k)}_t|^2] & =   E[|G^{(k)}_{t,t}|^2] \\
  &= k^2 \int_0^t
  E[|G^{(k-1)}_{t,s}|^2]\mathrm{d}\langle F^t \rangle_s \\
  &= k^2(k-1)^2
  \int_0^t \int_0^{s_1}
  E[|G^{(k-2)}_{t,s_2}|^2]\mathrm{d}\langle F^t \rangle_{s_2}
  \mathrm{d}\langle F^t \rangle_{s_1} \\
 & \leq (k!)^2 \langle F^t \rangle_t^k 
  =
   (k!)^2 \left(\frac{\eta^2}{4} \frac{|\tau^{-1}(t)|^{2H}}{\theta^{2H}}  \right)^k. 
  \end{split}
  \end{equation*}
  Note that $\tau^{-1}(t) \leq \tau^{-1}(1) = \theta$.
  Therefore, for sufficiently small $\theta$,
 \begin{equation*}
  \sum_{k=J}^\infty \theta^{Hk}
  \sum_{k=J}^\infty \frac{\theta^{Hk}}{(k!)^2}\int_0^1
  E[|G^{(k)}_t|^2]\mathrm{d}t
  \leq  \left( \frac{\eta^2}{4} \right)^J
\frac{\theta^{2HJ}}{(1 - \theta^H)(1-\theta^H\eta^2/4)},
  \end{equation*}
  which completes the proof. \hfill////

  Now we compute $a^{(i)}_\theta$,  $b_\theta$ and $c_\theta$ based on Lemma~\ref{lemrB}.
 The following lemmas follow from the results in Section~\ref{bb} by
  straightforward computations.
  \begin{lem}  \label{lemrB2}
  \begin{equation*}
  \begin{split}
  a^{(1)}_\theta(x) &= -H_3(x)\phi(x)\rho \eta \sqrt{\frac{H}{2}}\frac{\sigma_0(\theta)}{\theta^H}
\int_0^1 h_\theta(t)\int_0^t\frac{(\tau^{-1}(t) -
  \tau^{-1}(s))^{H-1/2}}{\sqrt{v_0(\tau^{-1}(s))}}\mathrm{d}s\mathrm{d}t\\
  &=-H_3(x)\phi(x)\rho \eta \sqrt{\frac{H}{2}}
  \\ & \hspace*{1cm} \times \frac{1}{\theta^H\sigma_0(\theta)^3}
  \int_0^{\theta}\exp\left\{-\frac{\eta^2}{8}t^{2H}\right\}\int_0^t
  (t-s)^{H-1/2}\sqrt{v_0(s)}\mathrm{d}s v_0(t)\mathrm{d}t \\
  & \sim
 -H_3(x)\phi(x)\frac{\rho \eta \sqrt{2H}}{2(H+1/2)(H+3/2)}.
  \end{split}
  \end{equation*}
  \end{lem}
  \begin{lem} \label{lemrB3}
  \begin{equation*}
  \begin{split}
  &a^{(2)}_\theta(x) = - H_2(x)\phi(x)\int_0^1 \frac{h_\theta(t)-1}{\theta^{2H}} \mathrm{d}t\\
  &
  -H_4(x)\phi(x)\rho^2 \frac{\eta^2 H}{4}
  \frac{\sigma_0(\theta)^2}{\theta^{2H}}\int_0^1 h_\theta(t)
  \left(\int_0^t
  \frac{(\tau^{-1}(t)-\tau^{-1}(s))^{H-1/2}}{\sqrt{v_0(\tau^{-1}(s))}}\mathrm{d}s\right)^2
  \mathrm{d}t \\
  & \sim H_2(x)\phi(x)\int_0^1 \frac{\eta^2}{8\theta^{2H}}|\tau^{-1}(t)|^{2H} \mathrm{d}t
  - H_4(x)\phi(x)\rho^2 \frac{\eta^2 H}{(2H+1)^2(2H+2)}.
  \end{split}
  \end{equation*}
  \end{lem}
  \begin{lem}  \label{lemrB4}
  \begin{equation*}
  \begin{split}
  a^{(3)}_\theta(x) &=
  -2H_2(x)\phi(x)\rho \eta \sqrt{\frac{H}{2}}\frac{\sigma_0(\theta)}{\theta^H}
\int_0^1\int_0^t\frac{(\tau^{-1}(t) -
  \tau^{-1}(s))^{H-1/2}}{\sqrt{v_0(\tau^{-1}(s))}}\mathrm{d}s\mathrm{d}t
  \\ & \sim -2H_2(x)\phi(x)\frac{\rho \eta \sqrt{2H}}{2(H+1/2)(H+3/2)}.
  \end{split}
  \end{equation*}
  \end{lem}
  \begin{lem}  \label{lemrB5}
  \begin{equation*}
  \begin{split}
  c_\theta(x) = & \  H_6(x)\phi(x) \rho^2
  \frac{\eta^2 H}{2} \frac{\sigma_0(\theta)^2}{\theta^{2H}}
  \left( \int_0^1 h_\theta(t) \int_0^t
  \frac{(\tau^{-1}(t)-\tau^{-1}(s))^{H-1/2}}{\sqrt{v_0(\tau^{-1}(s))}}
  \mathrm{d}s\mathrm{d}t\right)^2 \\
  & +  H_4(x)\phi(x)\rho^2
   \frac{\eta^2 H}{2} \frac{\sigma_0(\theta)^2}{\theta^{2H}}
   \int_0^1 h_\theta(t)^2\left( \int_0^t
  \frac{(\tau^{-1}(t)-\tau^{-1}(s))^{H-1/2}}{\sqrt{v_0(\tau^{-1}(s))}}
  \mathrm{d}s\right)^2\mathrm{d}t \\
  & +  H_4(x)\phi(x)\rho^2
   \eta^2 H \frac{\sigma_0(\theta)^2}{\theta^{2H}}
   \int_0^1 h_\theta(t) \int_0^t
  \frac{(\tau^{-1}(t)-\tau^{-1}(s))^{H-1/2}}{\sqrt{v_0(\tau^{-1}(s))}}
  \mathrm{d}s \\
  & \hspace*{4cm} \times
  \int_t^1 h_\theta(u)
  \frac{(\tau^{-1}(u)-\tau^{-1}(t))^{H-1/2}}{\sqrt{v_0(\tau^{-1}(t))}}
  \mathrm{d}u
  \mathrm{d}t \\
  & + H_4(x)\phi(x)
   \frac{\eta^2 H}{2} \frac{\sigma_0(\theta)^2}{\theta^{2H}}
   \int_0^1 h_\theta(t)^2\left( \int_s^1
  \frac{(\tau^{-1}(t)-\tau^{-1}(s))^{H-1/2}}{\sqrt{v_0(\tau^{-1}(s))}}
  \mathrm{d}t\right)^2\mathrm{d}s \\
  & +
  H_2(x)\phi(x)
   \frac{\eta^2 H}{2} \frac{\sigma_0(\theta)^2}{\theta^{2H}}
   \int_0^1 h_\theta(t)^2 \int_0^t
  \frac{(\tau^{-1}(t)-\tau^{-1}(s))^{2H-1}}{v_0(\tau^{-1}(s))}
  \mathrm{d}s\mathrm{d}t \\
   \sim & \  H_6(x)\phi(x)\rho^2
   \frac{\eta^2 H}{2(H+1/2)^2(H+3/2)^2}
   + H_4(x)\phi(x) \frac{2(1+\rho^2)\eta^2 H}{(2H+1)^2(2H+2)}\\
  &  + H_4(x)\phi(x) \frac{\rho^2\eta^2 H \beta(H+3/2,H+3/2)}{(H+1/2)^2}
 \\ & + H_2(x)\phi(x)\int_0^1 \frac{\eta^2}{4\theta^{2H}}|\tau^{-1}(t)|^{2H} \mathrm{d}t.
  \end{split}
  \end{equation*}
  \end{lem}

  \appendix
  \section{Conditional expectations of Wiener-It\^o integrals}\label{bb}
  Here we collect results on the conditional expectations of  Wiener-It\^o integrals that follow from Proposition~3 of Nualart et al~\cite{NUZ}.
  Let $x \in \mathbb{R}$ and $B$ be a standard Brownian motion ($B_0 = 0$).
  Let $f$ be a continuous function on
  \begin{equation*}
  \left\{ (s,t) \in (0,1)^2 ; s < t \right\}
  \end{equation*}
  with
  \begin{equation*}
  \int_0^1 \int_0^t |f(s,t)|^2\mathrm{d}s\mathrm{d}t < \infty.
  \end{equation*}
  \begin{lem}
  \begin{equation*}
  \begin{split}
  &  E\left[\int_0^1 \int_0^t f(s,t)\mathrm{d}B_s \mathrm{d}t \ \Big| \  B_1 = x
  \right] =
   H_1(x)\int_0^1 \int_0^t f(s,t)\mathrm{d}s\mathrm{d}t, \\
&  E\left[\int_0^1 \int_0^t f(s,t)\mathrm{d}B_s \mathrm{d}B_t  \ \Big| \  B_1 = x\right] =
   H_2(x)\int_0^1 \int_0^t f(s,t)\mathrm{d}s\mathrm{d}t, \\
  & E\left[\int_0^1\left( \int_0^t f(s,t)\mathrm{d}B_s\right)^2 \mathrm{d}B_t  \ \Big| \  B_1 = x\right]
   =
   H_3(x)\int_0^1 \left(\int_0^t f(s,t)\mathrm{d}s\right)^2
   \mathrm{d}t \\ & \hspace*{5.5cm} + H_1(x) \int_0^1 \int_0^tf(s,t)^2
   \mathrm{d}s\mathrm{d}t,\\
   & E\left[
   \int_0^1 \left(\int_s^1 f(s,t)\mathrm{d}B_t\right)^2 \mathrm{d}s
     \ \Big| \  B_1 = x \right]
   = H_2(x)   \int_0^1 \left(\int_s^1 f(s,t)\mathrm{d}t\right)^2 \mathrm{d}s
   \\
   & \hspace*{5.5cm} +  \int_0^1 \int_s^1f(s,t)^2
   \mathrm{d}t\mathrm{d}s.
   \end{split}
  \end{equation*}
  \end{lem}
  \begin{lem}
  \begin{equation*}
  \begin{split}
  & E\left[
\left(\int_0^1\int_0^t f(s,t)\mathrm{d}B_s \mathrm{d}B_t\right)^2
  \ \Big| \  B_1 = x \right]  - \int_0^1 \int_0^t f(s,t)^2\mathrm{d}s\mathrm{d}t\\ & = 
H_4(x) \left(\int_0^1\int_0^t f(s,t)\mathrm{d}s
  \mathrm{d}t\right)^2  +H_2(x)\int_0^1
  \left( \int_0^tf(s,t)\mathrm{d}s + \int_t^1 f(t,u)\mathrm{d}u\right)^2
  \mathrm{d}t.
  \end{split}
 \end{equation*}
  \end{lem}

\end{document}